\documentclass[aps, preprintnumbers,amsmath,amssymb,nofootinbib,notitlepage, onecolumn,superscriptaddress]{revtex4-1}
\pdfoutput=1
\usepackage{graphicx}
\usepackage[usenames,dvipsnames]{color}
\usepackage[bookmarks=true, bookmarksnumbered=true, colorlinks=true, urlcolor=darkblue, citecolor=darkgreen, linkcolor=darkred, breaklinks=true]{hyperref}
\definecolor{darkblue}{rgb}{0,0,0.5}
\definecolor{darkgreen}{rgb}{0.1,0,0.3}	
\definecolor{darkred}{rgb}{0.6,0,0}

\usepackage{multirow}		
\usepackage{mathrsfs} %To use multicolumn
\usepackage[T1]{fontenc}
\newcommand{\Neff}{N_{\rm eff}}
\newcommand{\ud}{\mathrm{d}}

\newcommand{\ms}{m_{s}}
\newcommand{\st}{\sin^2\!\theta\,}
\newcommand{\Hcal}{\mathcal{H}}

\begin{document}
\preprint{IFIC/14-53, FTUAM-14-32, IFT-UAM/CSIC-14-075}

\title{Revisiting cosmological bounds on  sterile neutrinos}

\author{Aaron C. Vincent}
\affiliation{Institute for Particle Physics Phenomenology (IPPP),\\ Department of Physics, Durham University, Durham DH1 3LE, UK.}
\affiliation{Instituto de F\'{\i}sica Corpuscular (IFIC)$,$
 CSIC-Universitat de Val\`encia$,$ \\  
 Apartado de Correos 22085$,$ E-46071 Valencia$,$ Spain}
 \author{Enrique Fern\'andez Mart\'inez}
 \affiliation{ Departamento and Instituto de F\'isica Te\'orica (IFT), UAM/CSIC,\\
C/ Nicol\'as Cabrera 13-15, Universidad Autonoma de Madrid, E-28049 Cantoblanco, Madrid, Spain}
\author{Pilar Hern\'andez}
 \affiliation{Instituto de F\'{\i}sica Corpuscular (IFIC)$,$
 CSIC-Universitat de Val\`encia$,$ \\  
 Apartado de Correos 22085$,$ E-46071 Valencia$,$ Spain}
 \author{Olga Mena}
   \affiliation{Instituto de F\'{\i}sica Corpuscular (IFIC)$,$
 CSIC-Universitat de Val\`encia$,$ \\  
 Apartado de Correos 22085$,$ E-46071 Valencia$,$ Spain}
\author{Massimiliano Lattanzi}
\affiliation{Dipartimento di Fisica e Science della Terra, Universit\`a di Ferrara and INFN,\\
sezione di Ferrara, Polo Scientifico e Tecnologico - Edficio C Via Saragat, 1, I-44122 Ferrara Italy}

 \begin{abstract}
We employ state-of-the art cosmological observables including supernova surveys and BAO information to provide  constraints on the mass and mixing angle of a non-resonantly produced sterile neutrino species, showing that cosmology can effectively rule out sterile neutrinos which decay between BBN and the present day. The decoupling of an additional heavy neutrino species can modify the time dependence of the Universe's expansion between BBN and recombination and, in extreme cases,  lead to an additional matter-dominated period; while this could naively lead to a younger Universe with a larger Hubble parameter,  it could later be compensated by the extra radiation expected in the form of neutrinos from sterile decay. However, recombination-era observables including the Cosmic Microwave Background (CMB), the shift parameter $R_{CMB}$ and the sound horizon $r_s$ from Baryon Acoustic Oscillations (BAO) severely constrain this scenario. We self-consistently include the full time-evolution of the coupled sterile neutrino and standard model sectors in an MCMC, showing that if decay occurs after BBN, the sterile neutrino is essentially bounded by the constraint $\sin^2\theta \lesssim 0.026 (m_s/\mathrm{eV})^{-2}$. 
 \end{abstract}

\maketitle
\section{Introduction}
Neutrino flavour change through the oscillation phenomenon is by now firmly established in solar, atmospheric, reactor and accelerator neutrinos and the determination of the neutrino mass differences and mixing angles that govern these oscillations has now entered the precision era, with only a couple of remaining unknown parameters and few percent accuracy in the rest (see \cite{GonzalezGarcia:2012sz} for an overview of the present status). This overwhelming evidence demands an extension of the Standard Model (SM) of particle physics able to accommodate the observed masses and mixings in the neutrino sector. 

Although several alternatives exist and the neutrino mass generation mechanism remains unknown, the simplest possibility is the extension of the SM particle content with right-handed neutrino fields $\nu_R$ in complete analogy with all other fermions of the theory. However, even if the addition of the $\nu_R$ only seems to make the lepton sector of the SM an exact copy of the quark sector with no significantly new phenomenology, the gauge singlet nature of the $\nu_R$ allows the existence of a Majorana mass term of the form $M \overline{\nu^c}_{R} \nu_R$, forbidden for any other fermion in the SM due to gauge invariance. The mass parameter $M$ introduces a completely new scale in the theory, unrelated to the electroweak scale and the Higgs mechanism, unlike all other elementary fermion masses in the SM. This Majorana mass term also violates lepton number $L$. Its running is therefore protected by lepton number symmetry and this scale will be stable under radiative corrections. The value of this new mass scale $M$ can thus take any possible value and remains an open question for experimental observations to address. Depending on the value of $M$, the corresponding phenomenology can be very different.

For very small $M\sim$~eV, extra sterile neutrinos could be present around the eV scale. These extra states could drive very short baseline oscillations and help to understand the experimental anomalies observed by some experiments like LSND, MiniBOONE and reactors. For a recent analysis of the oscillation data constraints on extra light eV species, see \cite{Abazajian:2012ys}. At the  $M\sim$~keV scale, extra sterile neutrinos can yield a viable warm dark matter candidate and could account for the dark matter (DM) component of the Universe --- this is the well-known Dodelson-Widrow scenario~\cite{Dodelson:1993je}. These DM particles decay to lighter neutrinos and photons and could be seen via X-ray searches~\cite{Abazajian:2001vt,Abazajian:2006yn, Kusenko:2009up}. Intriguingly, there is a recent claim for such an X-ray excess ~\cite{Bulbul:2014sua,Boyarsky:2014jta} in the form of a line at $E \sim 3.6$~keV, which could be explained by the decay of warm sterile neutrino Dark Matter with $M\simeq7.1$~keV~\cite{Ishida:2014dlp,Abazajian:2014gza}. For $M\sim$MeV--GeV the extra sterile states could affect the kinematics of weak decays or even induce flavour-changing processes in the lepton sector that would lead to characteristic signals~\cite{Atre:2009rg,Ruchayskiy:2011aa,Alonso:2012ji}. Finally, for $M > v$, the electroweak scale, the extra sterile states are heavy and can be integrated out at low energies. In this case, the Weinberg operator~\cite{Weinberg:1979sa} inducing neutrino masses emerges as the least suppressed low energy effect with inverse powers of $M$. In particular, light neutrino masses will be given by $m_\nu=m_D^t M^{-1} m_D$, where $m_D = Y v$ is the Dirac mass of the neutrinos obtained through their Yukawa couplings $Y$ and the vacuum expectation value of the Higgs $v$, as for any other fermion. In this case, the smallness of neutrino masses can be attributed to a hierarchy of scales between $M$ and $v$. This is known as the \textit{Seesaw mechanism} for the generation of $\nu$ masses~\cite{Minkowski:1977sc, Mohapatra:1979ia, Yanagida:1979as, GellMann:1980vs}. From a phenomenological point of view, these extra heavy steriles can also lead to deviation of universality in weak interactions as well as rare lepton flavour violating processes such as $\mu \rightarrow e \gamma$, $\mu \rightarrow 3e$ or $\mu \rightarrow e$ conversion in nuclei and in alternative channels involving the $\tau$ lepton~\cite{Langacker:1988up, Tommasini:1995ii, Antusch:2006vwa}. The Seesaw limit also offers the tantalizing possibility of explaining the observed baryon asymmetry of the Universe (BAU) through the leptogenesis mechanism~\cite{Fukugita:1986hr}.

Here we will show that extra sterile neutrinos with $M\sim$~keV--GeV can have a  significant impact on the evolution of the early Universe, leading to constraints on their mixing with charged leptons that can improve laboratory constraints by $\sim 10$ orders of magnitude in the mass range between few~MeV and few~GeV. Indeed, even a small interaction rate through the matrix element $U_{es}$ ensures that a small relic population of sterile neutrinos will be frozen out in the early Universe. If the mass and abundance of these particles are large enough, they might lead to an extra era of matter domination in the early Universe, which may alter the homogeneous expansion history, in addition to the matter perturbation spectrum and growth rate. In this work we focus on the former effects, showing that even a small matter contribution to the energy budget of the Universe during the radiation epoch can have a measurable effect on the observables quantified by precision cosmology experiments. 

Indeed, the introduction of an extra massive species in the early Universe can potentially lead to a faster expansion, and thus a younger Universe than the one we observe today. This is a well-known effect, and was used well before measurements of the CMB power spectrum to constrain the mass of the ordinary neutrino (see \textit{e.g.} \cite{Kolb:1990vq}) along with the existence of hypothetical extra neutrino species \cite{Sato1977,Gunn1978}. In the case of a heavy sterile species, this can be ``accidentally'' compensated by decay into a relativistic species, thus giving a longer-than-usual radiation period \cite{Dicus1979}, leading in turn to the correct Hubble parameter and luminosity-redshift relation for standard candles such as supernovae. However, we will show that early-time observables, especially the sound horizon measured by BAO, severely limit this scenario, essentially restricting the presence of any heavy ($\gg$ keV) sterile neutrino that would decay after nucleosynthesis;  these bounds are similar to limits placed on a decaying dark matter particle into invisible radiation \cite{Ichiki:2004vi, Gong:2008gi,GonzalezGarcia:2012yq,Lattanzi:2013uza,Audren:2014bca}. The typical approach is to use constraints on matter and/or radiation obtained within a $\Lambda$CDM model. We will illustrate these naive bounds in Sec.~\ref{sec:sterileTheory} before describing our full self-consistent cosmological approach, which involves the simultaneous solution of the evolution equations for each cosmological component. Our aim is therefore to build a consistent framework to study the combined effect of thermal production, freeze-out and decay of a sterile neutrino species on the most up-to-date background cosmological observables including the Planck, BOSS and SDSS experiments, along with the SNLS supernova survey and the Hubble Space Telescope. We will show that these cosmological constraints severely limit the available parameter space in this scenario, effectively restricting the sterile neutrino mixing with the SM to be less than $\sin^2 \theta < 10^{-4}$ for small masses $m_s  \sim $ eV down to  $\sin^2 \theta < 10^{-17}$ for $m_s  = 100$\,MeV. 

We begin in Sec.~\ref{sec:sterileTheory} with a summary of the thermal production and freezeout mechanism through mixing for a relic sterile neutrino species, along with its decay rate, and general features of its effect on standard cosmology. This is followed in Sec.~\ref{sec:cosmo} by a more detailed description of the modified cosmological evolution due to such an extra species, which we implement into an MCMC algorithm. In Sec.~\ref{sec:methods} we present the cosmological observables and datasets used to obtain our constraints, which are shown in Sec.~\ref{sec:results}. We conclude with a comparison with other known constraints at the end of Sec.~\ref{sec:results}, and we conclude in Sec.~\ref{sec:discussion}.

\section{Decoupling and abundance}
\label{sec:sterileTheory}
We consider here a model with one extra sterile Majorana species $\nu_s$ of mass $m_s$,  that can mix with the three standard neutrinos: 
we define $\sin^2 \theta \equiv \sum_{\alpha=e,\mu,\tau} |U_{\alpha s}|^2$. Both the production rate of sterile neutrinos in the early Universe and their decay depend very sensitively on the mass and mixing. We will focus on the constraints on sterile species that decay after active neutrino decoupling and therefore have masses below 1 GeV (cosmology bounds on heavier species are expected to be significantly weaker). To remain as model-independent as possible, we will not include correlations between masses, mixings and light neutrino masses, that would be generic in minimal models where the sterile mass results from a low-scale seesaw \cite{Hernandez:2014fha}. 

\begin{figure}[h]
\begin{tabular}{c c }
\includegraphics[width=.45\textwidth]{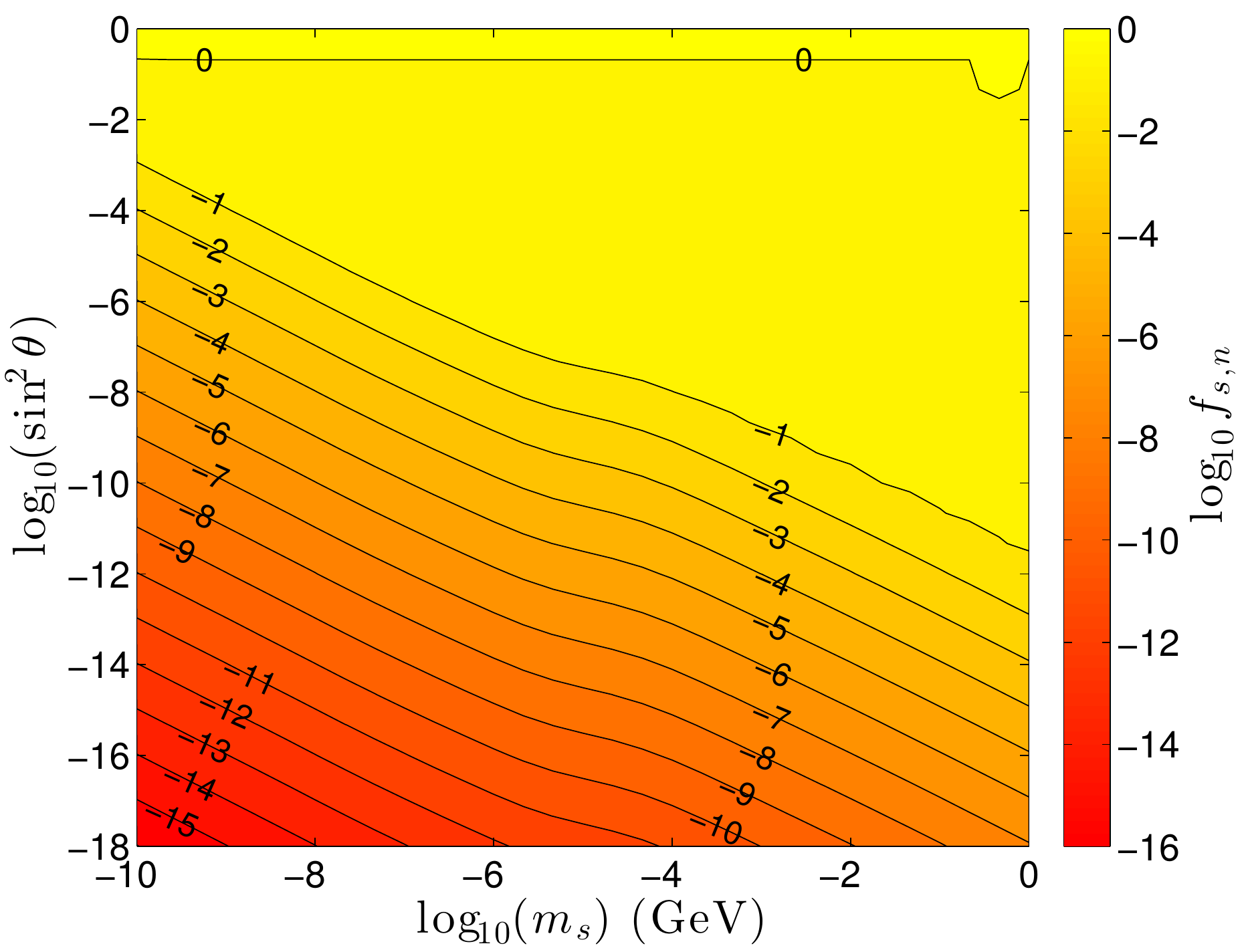}  &\includegraphics[width=.45\textwidth]{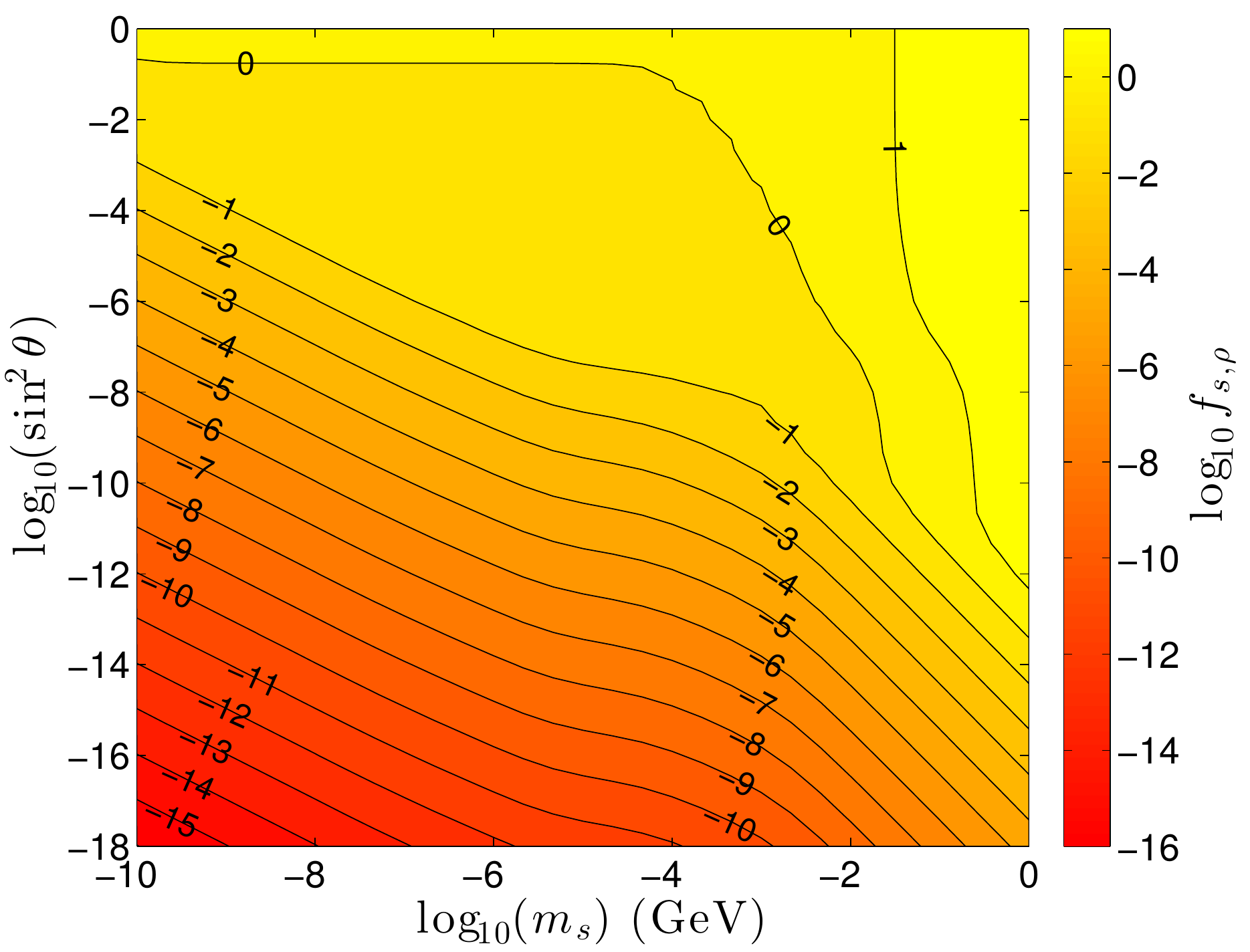} 
\end{tabular}
\caption{``Suppression factors'' $f_{s,n}$ (left) and $f_{s,\rho}$ (right) evaluated at $T = 1\,{\rm MeV}$ as defined in Eqs.~(\ref{ninit}-\ref{rhoinit}), as a function of the mass $m_s$ and mixing angle $\sin^2\theta$; these are computed using the standard methods of Ref.~\cite{Dolgov:2002wy}. Labels correspond to isocontour values of $\log_{10} f_{s,i}$. Although we only show the case of mixing with electron neutrinos, other channels differ by only a few percent, which is not visible on this scale.}
\label{fig:cms}
\end{figure}

It is well-known \cite{Barbieri:1989ti,Barbieri:1990vx,Kainulainen:1990ds} that sterile neutrinos with masses below a GeV are effectively produced in the early Universe via mixing, which is  however strongly modified in the thermal plasma
 \cite{Notzold:1987ik}. In the absence of primordial lepton asymmetries,  the rate of production is maximal at a temperature  $T_{\rm max}\simeq 100$\,MeV$(m_s/$keV)$^{1/3}$, although the coefficient is slightly dependent on the flavour if each flavour mixes differently. Note that $T_{\rm max} \gg m_s$ in the 
range we are interested in. Provided that the rate of production at
this temperature is larger than the Hubble rate,  sterile neutrinos will reach thermal equilibrium and 
decouple at a temperature which is related to the decoupling temperature of the active neutrinos,  $T_{dec,\nu} \simeq 2.3$\, MeV,  by  
\begin{equation}
T_{dec} = T_{dec,\nu} \sin^{-2/3} \theta.
\label{Tdec}
\end{equation}
If the sterile neutrino production rate never reaches the Hubble rate, their abundance will be suppressed with respect to the thermal abundance. Using standard methods \cite{Dolgov:2002wy}, one may relate the sterile neutrino density to that of a single, active neutrino species as a function of temperature\footnote{For $m_s \geq$\,keV, the production occurs at or above the QCD phase transition and there is a significant hadronic uncertainty in the active neutrino interaction rate \cite{Abazajian:2002yz,Abazajian:2005gj,Asaka:2006nq}. As in \cite{Hernandez:2014fha}, we included only leptonic contributions to the active neutrino interaction rate. }. We thus define the ``suppression factors'' in the number density,  $f_{s,n}(T) \equiv n_s(T)/n_{1 \nu a}(T)$ and in the energy density, $f_{s,\rho}(T) \equiv \rho_s(T)/\rho_{1 \nu a}(T)$, where the subscript ``$1\nu a$'' refers to a single active neutrino species, rather than all three\footnote{These factors and, consequently, the cosmological constraints that we derive here, will change in non standard cosmologies, as for instance, in low reheating temperature scenarios~\cite{Gelmini:2004ah}.}. The values $f_{s,i}$ thus relate the asymptotic solution to the Boltzmann equations during the production of the sterile neutino species, to what is expected from a standard thermal relic, thus allowing for a more intuitive parametrization of the cosmological abundances. The sterile neutrino number and energy densities before decays are then:
\begin{eqnarray}
n_s(T > T_{\mathrm{decay}}) &=& f_{s,n}(T) \frac{3}{2}\frac{\zeta(3)}{\pi^2}T^3, \label{ninit}\\
\rho_s(T > T_{\mathrm{decay}})&=& f_{s,\rho}(T) \frac{7}{4}\frac{\pi^2}{30}T^4.
\label{rhoinit}
\end{eqnarray}
Note that the ``suppression factor'' can be larger than one: this is the case if the neutrinos become non-relativistic between $T_{\rm max}$ and the temperature at which $f_s(T)$ is evaluated, resulting in less dilution than the active neutrino species. We have computed the suppression factors following the method of \cite{Hernandez:2014fha}) at $T = 1$\,MeV, where we begin our cosmological evolution. The results are shown in Fig.~\ref{fig:cms} for mixing of the sterile with electron neutrinos. Other flavour channels yield only a few percent difference in the values of $f_{s}$.
Note that there is no Boltzmann suppression because the sterile neutrino falls out of kinetic equilibrium while relativistic for this range of parameters \cite{Hernandez:2014fha}. 

Sterile neutrino decay can occur via mixing in various channels, depending on the mass. These are detailed in Refs. \cite{ Gorbunov:2007ak,Fuller2011}. The main effect of such channels opening is a reduction in the sterile neutrino lifetime. This depends strongly on the ratio of couplings to electron, mu and tau neutrinos. For our analyses we take one fiducial case, where $\nu_s$ couples only to $\nu_e$. 

There are further effects that arise from decay to heavy species: electron production after recombination can slightly reionize the Universe, rescattering CMB photons. Decay before recombination can heat the plasma, leading to a larger radiation. We illustrate these effects separately, at then end of Sec. \ref{sec:results}.

The decay rate into three active neutrinos is given by:
\begin{equation}
\Gamma_s = \frac{G_F^2}{192\pi^3}\st \ms^5.
\end{equation}
In Fig.~\ref{H0clean}, we show the effective decay times $t = \tau_s \equiv 1/\Gamma_s$ for $\tau_s=t_{BBN} \simeq 10$\,s, $\tau_s=t_{CMB}$ (recombination),  and the present time, $\tau_s = t_0 = 13.5$\,Gyr.  Due to decay, the time evolution of the sterile and active neutrinos are coupled. Many studies have considered BBN bounds on this model, assuming full thermalisation of the sterile species; a detailed recent analysis  can be found in \cite{Ruchayskiy2012}.  We will consider the parameter space allowed by those analyses where decay occurs after BBN. We thus follow the background evolution after BBN, taking into account both the possible non-thermal distribution of the sterile neutrinos and their decay. In this case, the evolution of the density of sterile neutrinos and their decay products decouple from  
that of radiation. Before turning to the full cosmological model described in Sec.~\ref{sec:cosmo}, it is instructive to derive some approximate bounds on this scenario.

Cosmological bounds on the sterile parameters \cite{Kolb:1990vq}  can be estimated in a rather naive way as follows. If the sterile neutrinos have a lifetime 
longer than the age of the Universe, they contribute to the energy density as cold, warm or hot dark matter depending on $m_s$~\cite{Abazajian:2001nj}. Requiring that their contribution to 
$\Omega$ is smaller than $\Omega_m$ gives the bound
\begin{eqnarray}
\Omega_s h^2 \simeq 10^{-2} f_{s,n} m_s(eV) \leq \Omega_m h^2 \simeq 0.12,  
\label{OmegaBound}
\end{eqnarray} 
which corresponds to the thick red line in Fig.~\ref{H0clean}. Note that this would correspond to all of the DM being the sterile neutrino component, therefore we cut the line at $\tau_s = t_0$ as a requirement that the dark matter still be present in the Universe today. 

Finally if decay occurs before recombination, the sterile component will be constrained from its contribution to dark radiation at recombination, since the sterile neutrino decay 
will produce an active neutrino component with a very non-thermal distribution (the distribution is determined from the decay kinematics of the massive sterile neutrino and not by the temperature). If all the energy density is transferred to dark radiation at  the temperature of decay, $T_{decay}$, this extra radiation component can be estimated as 
\begin{eqnarray}
\Delta N_{\rm eff} &\simeq &\frac{180}{ 7 \pi^4 } \zeta(3)  f_{s,n} \frac{m_s }{ T_{decay}}  \\
&\simeq& 1.3 \times 10^8 \left(\frac{m_s}{\rm GeV}\right)^{-1/2} \sin\theta \nonumber 
\label{deltaNeff}
\end{eqnarray}
The thick orange line in Fig.~\ref{H0clean} corresponds to $\Delta N_{\rm eff} = 1$ using this naive approach; the value of $\Delta N_{\rm eff}$ increases as one moves above this line.

Naively we would expect that, for $\Gamma_s^{-1} \geq t_{BBN}$,  the region limited by the red and orange lines is excluded. 
These bounds are naive because they use constraints on $\Omega_m$ or $\Delta N_{\rm eff}$ that are derived within a $\Lambda$CDM model. Our objective is 
to derive robust bounds based on background cosmology observables obtained self-consistently within this $\Lambda$CDM-sterile cosmology.

%%%%%%%%%%%%%%%%%

\section{Cosmological evolution}
\label{sec:cosmo}
The unperturbed, homogeneous and isotropic background is written as a FRW metric: $ds^2 = -dt^2 + a^2(t) d\vec x ^2$, where the scale factor $a(t)$ parametrizes the expansion of the Universe and is the reciprocal of the redshift: $a(t) = 1/(1+z)$. The evolution of the heavy sterile neutrino energy density and its decay products can be written in the following simple way:
\begin{eqnarray}
y_s'(x) &=& -3(1+w_s(x))y_s(x) - \frac{\gamma_s}{\Hcal(x)}y_s(x); \label{yheavy} \\
y_\nu'(x) &=& -4y_\nu(x) + \frac{\gamma_s}{\Hcal(x)}y_s(x); \label{yproducts}
\label{evoeqs}
\end{eqnarray}
where the dimensionless time $x \equiv \ln a(t)$ is the log of the scale factor, and the density $y_j \equiv \rho_j/ \tilde \rho$. $\tilde \rho$ is an arbitrary normalization scale. 

The matter, radiation and cosmological constant ($y_m$, $y_{rad}$ and $y_\Lambda$) are described in Appendix \ref{cosmoappendix} and combine with (\ref{yheavy}-\ref{yproducts}), in the dimensionless Friedmann equation:
\begin{equation}
\Hcal^2(x) = \sum_j y_j
\label{friedmann}
\end{equation}
 $\Hcal$ is related to the physical Hubble parameter via $\Hcal = H/\tilde H$, with $\tilde H = 8\pi G \tilde \rho/3$. Finally, $\gamma_s$ is the dimensionless sterile neutrino decay rate $\gamma_s \equiv {\Gamma_s}/{\tilde H}$.
 
One final subtlety lies in the equation of state parameter $w_s$ of the sterile neutrino. Unless it is very heavy, $\nu_s$ will be relativistic when it decouples. This means that the equation of state $w_s(T) = {P_s(T)}/{\rho_s(T)}$ must be tracked during the evolution. The phase space distribution is frozen in at the decoupling temperature $T_{\rm dec}$, and the energy density is therefore:
\begin{eqnarray}
\rho_s(T) &=& \frac{4\pi g}{(2\pi)^3}\int_0^\infty\frac{p^2(T) \sqrt{p^2(T)+m^2} }{e^{\sqrt{p_0^2+m^2}/T_{dec}}+1}\ud p  \nonumber \\
&=& \frac{g}{2\pi^2}T^4 \int_0^\infty \frac{u^2\sqrt{u^2+(m_s/T)^2}}{e^{\sqrt{u^2 + (m_s/T_{\rm dec})^2}}+1} \ud u,
\end{eqnarray}
where $p_0$ is the momentum at the time of decoupling. The pressure is:
\begin{eqnarray}
P_s(T) &=&  \frac{4\pi g}{(2\pi)^3}\int_0^\infty\frac{p^2(T)}{3 E(T) }\frac{p^2(T) }{e^{\sqrt{p_0^2+\ms^2}/T_{dec}}+1}\ud p  \nonumber \\
 &=& \frac{g}{6\pi^2}T^4 \int_0^\infty \frac{u^2}{\sqrt{u^2+(m_s/T)^2}} \frac{u^2}{e^{\sqrt{u^2+ (\ms/T_{\rm dec})^2}}+ 1}\ud u.
\end{eqnarray}
Equations (\ref{ystandard}), (\ref{yheavy})-(\ref{yproducts}) can be numerically solved from the time of sterile neutrino decoupling $a = a_{dec}$ until today, $a = 1$. In practice, we begin  the evolution at $T = 1$\, MeV, shortly after active neutrino decoupling. This corresponds to a scale factor of $a(T= 1\,\mathrm{MeV}) = 1.7 \times 10^{-10}$, and allows us to use the well-defined initial condition for $\rho_s(T=1\,\mathrm{MeV})$ defined in (\ref{rhoinit}).  Since direct measurements of the CMB temperature fix the radiation density, the model inputs are the sterile mass and mixing ($\ms$, $\st$), the initial matter density $y_m$ and the dark energy density $y_\Lambda$. For every combination of these four parameters, one obtains a Universe with a given age (or expansion rate, parametrized by $H_0 \equiv \tilde H \Hcal(0)$) and composition (parametrized by $\Omega_i \equiv y_i(0)/\Hcal(0)$). 
Some examples are shown in Figure \ref{fish} of the following section, where we will use the  approach described above in order to constrain the sterile neutrino  parameter space.

%%%%%%%%%%%%%%%%%%%%

\section{New constraints from background cosmology}
\label{sec:constraints}
\subsection{Method and cosmological measurements}
\label{sec:methods}
 We perform an MCMC scan over the parameter space $m_s = \left[10^{-10}, 1\right]$ GeV, and $\st = \left[10^{-18}, 10^0\right]$. For consistency, we also vary the density of matter $\Omega_m$ and of dark energy $\Omega_\Lambda$. Since the observable quantities that we are interested in pertain only to the background evolution, other quantities that normally go into $\Lambda$CDM analyses, namely $A_s, n_s, \tau_{reio}$ do not enter into our calculation, and are therefore not part of our analysis. Changing $\Omega_b$ by the amount allowed by BBN constraints on helium and deuterium production would induce negligible changes in the recombination redshift, and therefore in the shift parameter. We perform two separate MCMC analyses: one with decays to heavy species included, assuming coupling to $\nu_e$, and one with only decays to $3\nu$. Our main result includes decays to $\pi^0 \nu$, $\pi^\pm e^\mp$, $\pi^\pm \mu^\mp$, $K^\pm e^\mp$ and the three-body decays to $\nu e^+ e^-$ and $\nu \mu^+ \mu^-$.
 
   In every case, we fix the present-day CMB temperature to the observed one, defining the scale factor today as $a_0 \equiv a(T_\gamma = T_{\rm CMB}) = 1$. For our MCMC analyses, we consider the cosmological measurements related exclusively to the Universe's background expansion history: measurements of the Hubble constant, Supernovae Ia luminosity distances, the CMB shift parameter, as well as measurements of the Baryon Acoustic Oscillation scale. We briefly describe all these measurements in what follows. Concerning the value of the Hubble constant $H_0$, we apply a gaussian prior of $H_0=73.8\pm 2.4$ from the Hubble Space Telescope~\cite{Riess:2011yx}. Type  Ia Supernovae (SNIa) luminosity distance data is also sensitive to the Universe's expansion rate at low redshift, via the distance modulus $\mu$:
\begin{equation}
\mu \equiv 5 \log_{10}\left(\frac{d_L(z)}{\textrm{Mpc}}\right) +25~,
\label{eq:dL}
\end{equation}
where $d_L(z)$ represents the luminosity distance $d_L(z) = c (1+z) \int_0^z 1/H(z') dz'$. We use the distance moduli from the 3-year Supernova Legacy Survey (SNLS) data set \cite{Conley:2011ku}, which consists of 115 SNIa with redshifts up to $z\sim 1$.

Cosmic microwave background temperature anisotropy measurements from the Planck experiment~\cite{Ade:2013zuv} are  included via the CMB shift parameter $R_{CMB}$, defined as
\begin{equation}
R_{CMB}= \sqrt{\Omega_m H_0^2} \int_0^{z_{\rm rec}}\frac{dz}{H(z)}~,
\label{eq:shift}
\end{equation}
where $z_{\rm rec}$ is the redshift of recombination.  $R_{CMB}$ is the least model-dependent quantity extracted from the CMB power spectrum, as it is independent of the measured value of $H_0$.  It is included in our analyses via a gaussian prior: $R_{CMB}= 1.7407\pm0.0094$~\cite{Wang:2013mha}. We note however that, while in standard cosmologies the value of the recombination redshift may be easy computed via the empirical parameterisation from Ref.~\cite{Hu:1995en}, in non-standard scenarios such as the one explored here the value of the recombination redshift must be computed numerically at each step of the MCMC analysis. Therefore, $z_{\rm rec}$ is computed for each possible combination of ($m_s, \st, \Omega_m$ and $\Omega_\Lambda$) by means of the RECFAST software~\cite{Seager:1999bc,Seager:1999km,Wong:2007ym,Scott:2009sz}, which provides the precise free electron fraction evolution as a function of the redshift $x_e(z)$. The recombination redshift $z_{\rm rec}$ is defined as the redshift at which the optical depth $\tau(z)$ is equal to one:
\begin{equation}
\tau(z_{rec}) = \int^{z_{\rm rec}}_{0} dz \frac{d\eta}{da} x_e(z)\sigma_T \equiv 1~,
\label{eq:zeq}
\end{equation}
\noindent 
where $\eta$ is the conformal time and the $\sigma_T$ is the Thomson cross section.

Before the recombination epoch, the competition between gravity and radiation pressure in the photon-baryon fluid leads to oscillations in the plasma which propagate as acoustic waves known as Baryon Acoustic Oscillations (BAO). At recombination ($z \sim 1100$), the photons decouple from the baryons and start to free stream, whereas the pressure waves remain frozen. As a result, baryons accumulate at a fixed distance from the original over-density, equal to the sound horizon length at the decoupling time. The result is a peak in the mass correlation function at this distance, providing, a standard ruler (the BAO scale) which can be measured at various redshifts using the clustering distribution of galaxies inferred from galaxy surveys.  We consider the WiggleZ Survey measurements of the BAO acoustic peak at three different redshifts $z=0.44, 0.6$ and $0.73$~\cite{Blake:2011en}, in terms of $d(z)\equiv r_s(z_{drag})/D_V(z)$, where $r_s$ is the sound horizon at the drag epoch ($z_{drag}$) and $D_V(z)$ represents the spherically averaged clustering statistics
\begin{equation}
D_V(z)= \left((1+z)^2 D_A(z)^2 \frac{cz}{H(z)} \right)^{1/3}~;
\label{eq:DV}
\end{equation}
 $D_A(z)$ is the physical angular diameter distance, $D_A(z)=d_L(z)/(1+z)^2$, see Eq.~(\ref{eq:dL}). The sound horizon $r_s(z_{drag})$ is defined as the comoving distance that a wave can travel from the beginning of the Universe until the drag period
\begin{equation}
r_s(z_{drag})= \int^{\eta(z)}_0 d\eta \ c_s (1+z)~,
\label{eq:sound}
\end{equation}
where $c_s=1/\sqrt{3(1+R)}$ is the sound speed, and $R\equiv 3\rho_b/4 \rho_\gamma$. The drag epoch corresponds to the redshift at which the \textit{drag} optical depth $\tau_d$ is equal to one:
\begin{equation}
\tau_d(z_{drag})  =\int^{z_{drag}}_{0} dz \frac{d\eta}{da}\frac{x_e(z)\sigma_T}{R} \equiv1~.
\label{eq:zdrag}
\end{equation}
For non-standard cosmologies, $z_{drag}$ and $r_s(z_{drag})$ must again be recomputed at each point in the parameter space \cite{Hamann:2010pw}. Once more, we compute the free electron fraction $x_e(z)$, with the RECFAST code before solving Eqs.~(\ref{eq:sound}) and (\ref{eq:zdrag}) numerically.

In addition, depending on whether the BAO scale is measured along or across the line of sight, it provides a redshift-dependent measurement of the Hubble parameter or of the angular diameter distance, respectively. We use here the BAO measurements from the Data Release 11 (DR11) of the Baryon Oscillation Spectroscopic Survey (BOSS)~\cite{Dawson:2012va}, one of the four surveys of the Sloan Digital Sky Survey III (SDSSIII) experiment~\cite{Eisenstein:2011sa}. The Lyman alpha forest absorption BAO features observed in the BOSS DR11 quasars catalog, provides, at an effective redshift of $z=2.36$, $c/(H(z=2.36) r_s( z_{drag})=9.0\pm 0.3$ and $D_A(z=2.36)/ r_s( z_{drag})=10.8 \pm 0.4$~Mpc~\cite{Font-Ribera:2014wya}. On the other hand, BOSS DR11 measurements of the BAO signal in the clustering of galaxies provide, at an effective redshift of $z = 0.57$, $D_A(z=0.57)=1421\pm 20$ Mpc $\times (r_s(z_{drag})/r_{s,fid}) $ and $H(z=0.57)= 96.8\pm 3.4$ km/s/Mpc $\times (r_s(z_{drag})/r_{s,fid})$, where $r_{s,fid}=149.28$~Mpc. 

\subsection{Results}
\label{sec:results}
We have considered two possible data combinations, which measure two qualitatively different effects on the observables. We first perform a MCMC using the HST prior and the SNLS luminosity-distance measurements only; effectively constraining the current age of the Universe and its post-recombination evolution. We then compute a second and more complete run which probes background cosmology at many epochs, combining the two former data sets with the CMB shift parameter $R_{CMB}$ and the BAO constraints described in Sec.~\ref{sec:methods}. 
\begin{figure}[h]
\includegraphics[width=.7\textwidth]{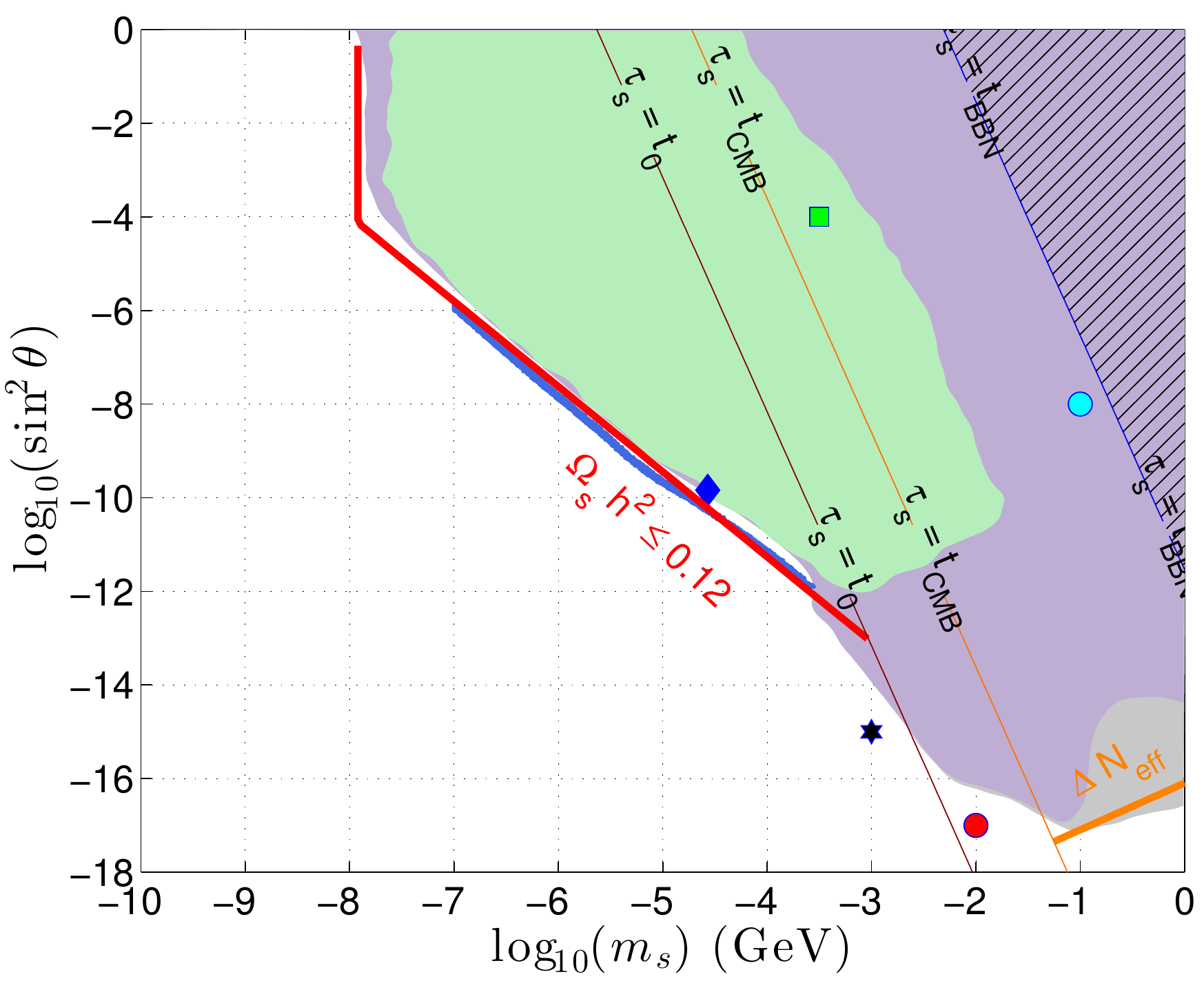}
\caption{Excluded regions from the results of our Markov Chain Monte Carlo analysis, using the cosmological data described in Sec.~\ref{sec:methods}. Light green region: $H_0$ only. Purple region: full data set including BAO. Particles that decay before BBN (hashed region) cannot be constrained with this method. The lighter region in the lower-right illustrates the exclusion based on decays to $3 \nu$ only, whereas the purple exclusion region includes decays to heavier species, based on the branching ratios given by \cite{Gorbunov:2007ak}. The red and orange lines correspond to the naive bounds from Eqs.~(\ref{OmegaBound}) and (\ref{deltaNeff}), explained in Sec.~\ref{sec:sterileTheory}. They respectively correspond to a limit on the total abundance of DM $\Omega_s h^2 \lesssim 0.12$, and to the requirement that $\Delta \Neff \lesssim 1$. The darker blue line represents the region for which the sterile neutrino can act as the cosmological dark matter which is favoured by our MCMC analysis. Each symbol represents one of the cosmological scenarios shown in Fig.~\ref{fish}. Finally, the thin lines are lines of equal sterile neutrino lifetimes, corresponding to decays that occur at present ($t_0$ = 13.5 Gyr), at recombination ($ t_{CMB} =$ 380 000 yr) and at BBN ($t_{BBN} = 10$ s). }
\label{H0clean}
\end{figure}
\begin{figure}[p]
\begin{tabular}{c@{\hspace{0.04\textwidth}}c}
\multicolumn{2}{c}{\includegraphics[width = 0.4\textwidth]{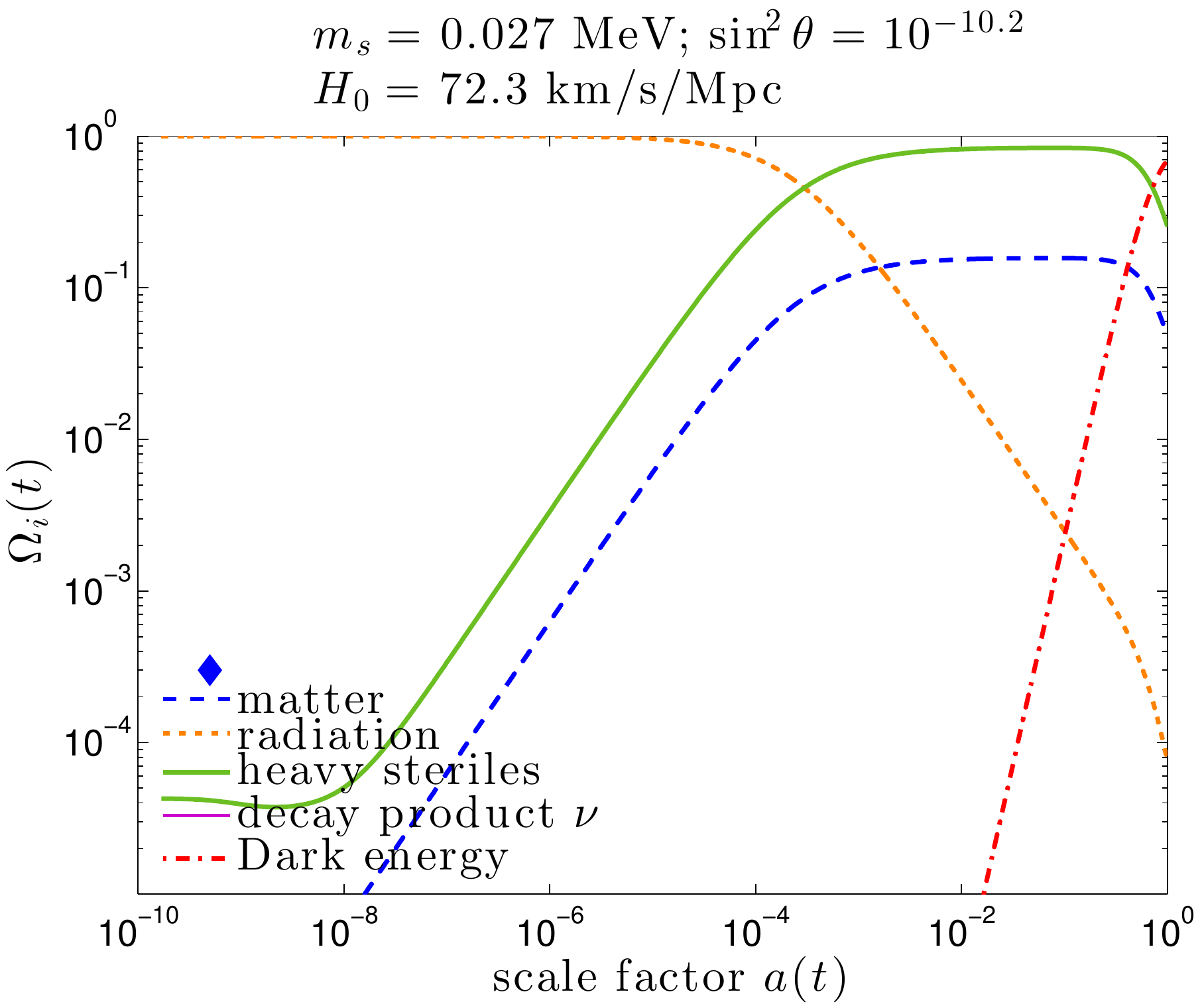}} \\
\includegraphics[width=.4\textwidth]{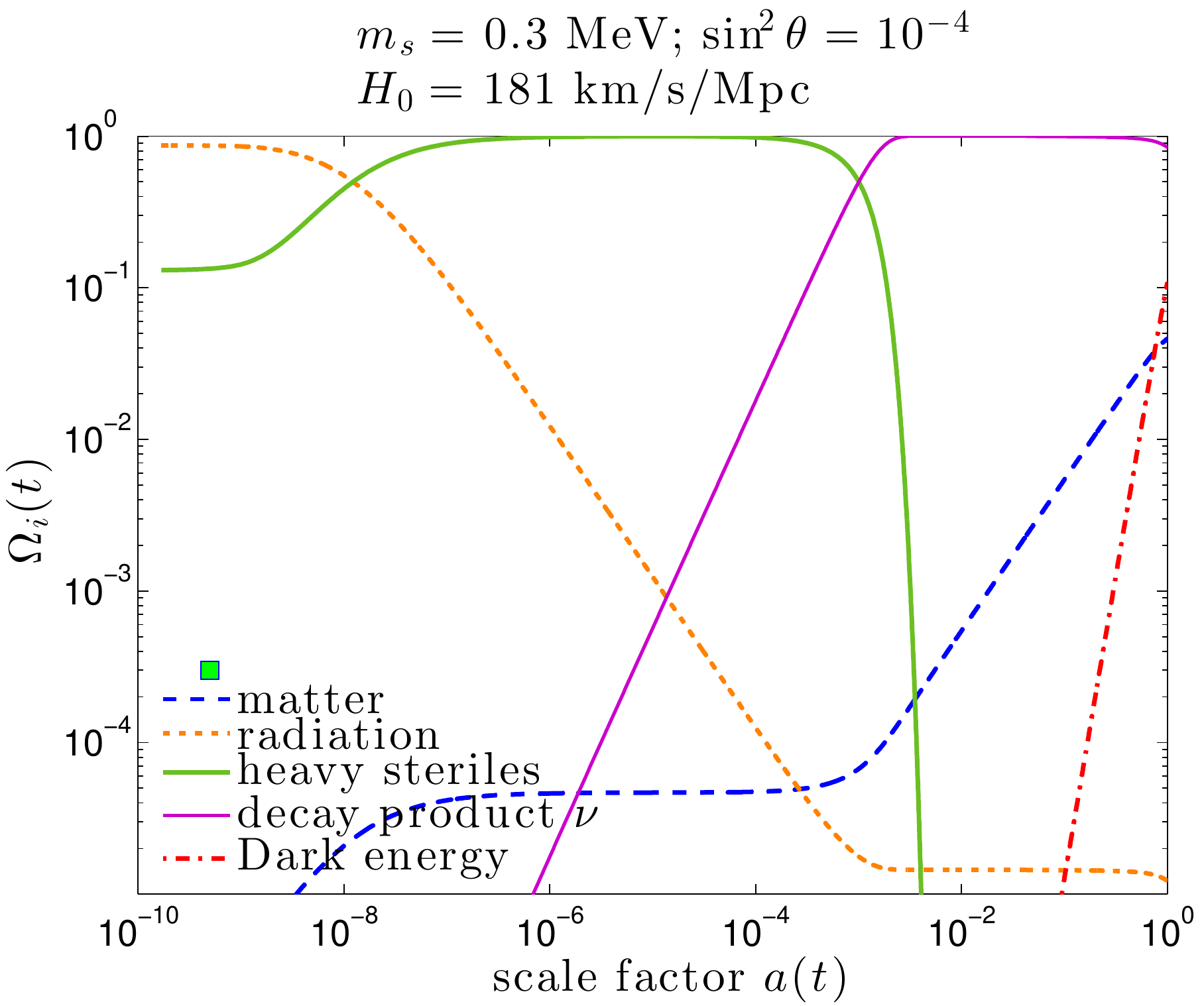} & \includegraphics[width=.4\textwidth]{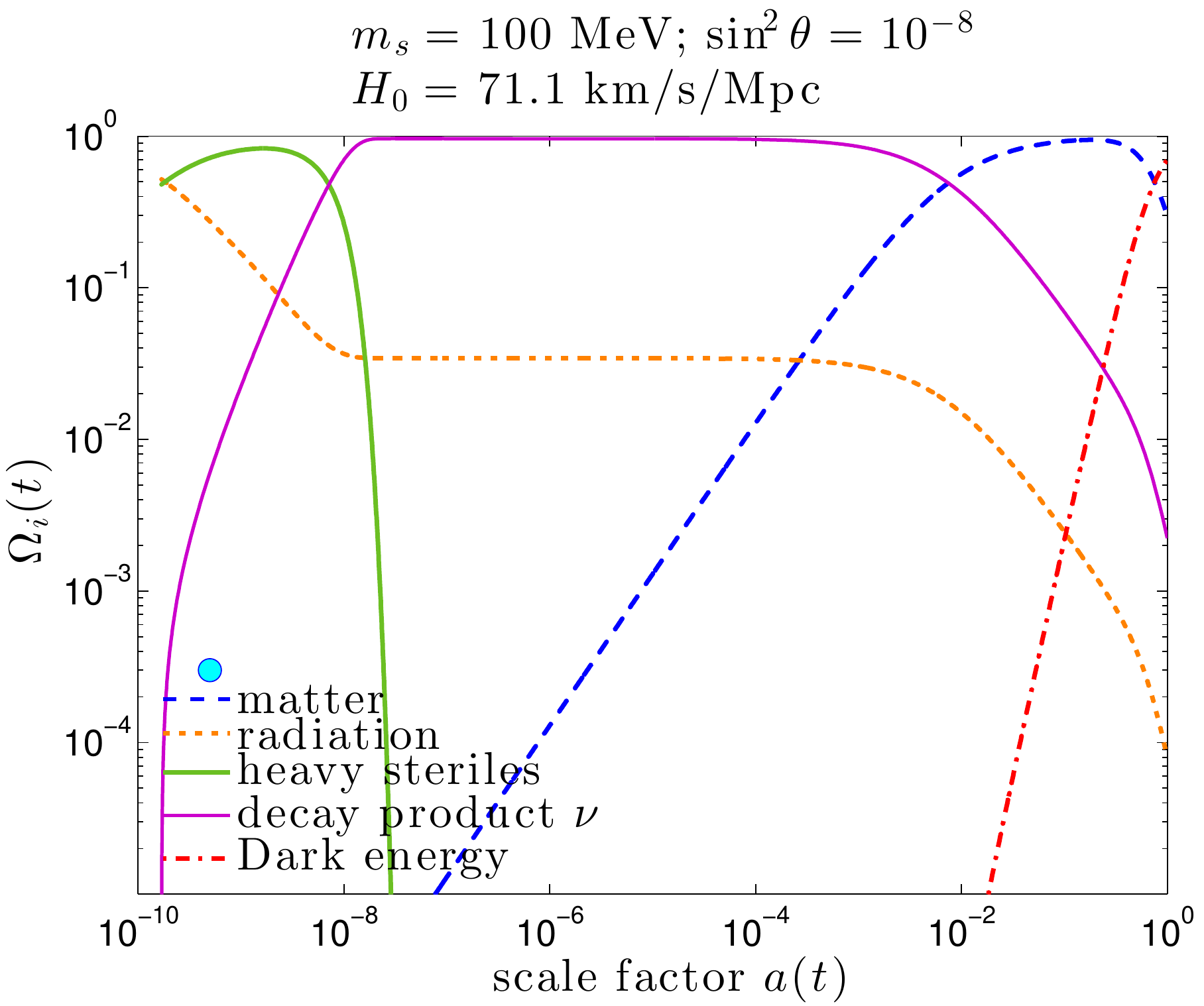}\\
\includegraphics[width=.4\textwidth]{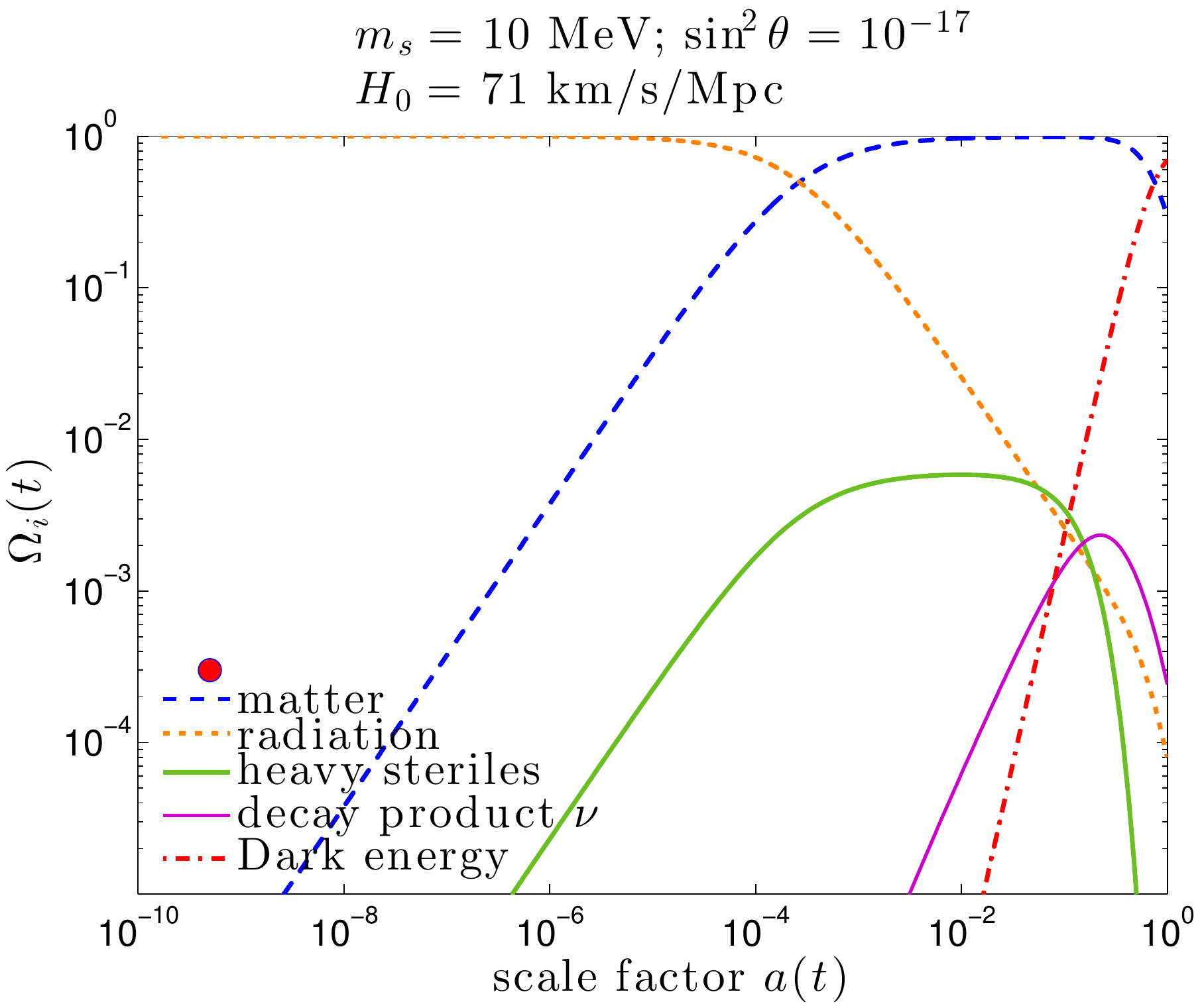} & \includegraphics[width=.4\textwidth]{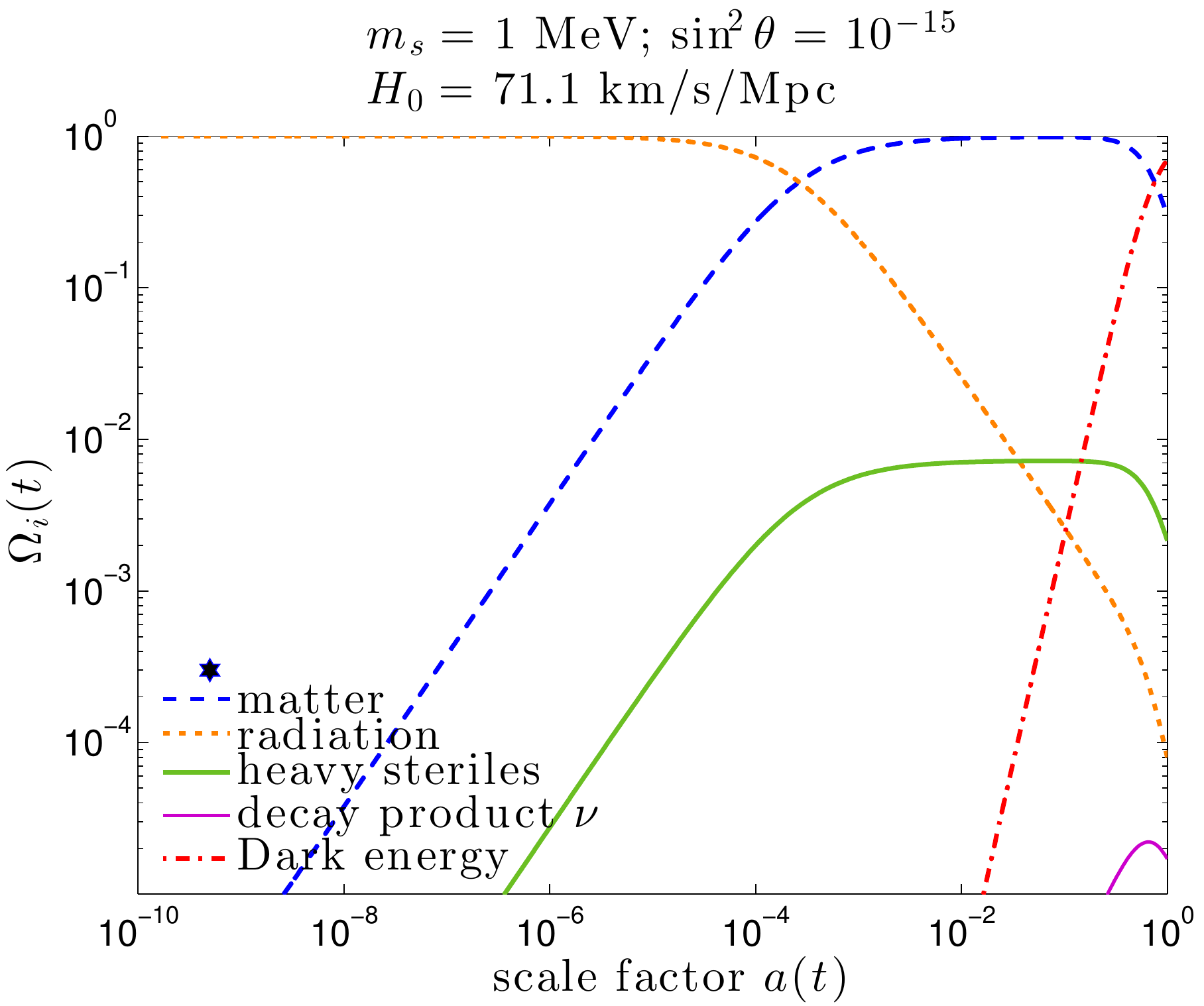}
\end{tabular}
\caption{Relative contribution $\Omega_i \equiv \rho_i/\rho_c$ of each component for five different examples of the sterile neutrino mass and coupling combinations, as a function of the scale factor $a(t) = 1/(1+z(t))$. $a(t_0) = 1$ corresponds to the present ($z = 0$). In every case the symbol above the legend shows the scenario's position in Fig.~\ref{H0clean}. The upper panel shows the case in which the heavy neutrino, with a mass is in the keV region and lifetime larger than the age of the Universe, represents the dominant component to the dark matter fluid; in this case the ``matter'' line refers only to baryons. The middle plots correspond to cases which are in strong tension with the cosmological data sets used here. The left-hand middle scenario lies in the green region of Fig.~\ref{H0clean} and yields a matter-dominated period that alters the late-time expansion rate enough to be ruled out by observations of $H_0$. The right-hand middle figure, which lies in the purple region of Fig. \ref{H0clean}, yields the correct present-day cosmology; however, the rapid expansion induced by a short matter-dominated period does not reproduce the correct sound horizon at recombination $r_s(z_{drag})$ and is thus ruled out by BAO observations. The lower two panels show two examples allowed by the data considered here in which a significant population of sterile neutrinos decay after the recombination period. In the left plot, the sterile states have decayed to active neutrinos before the current epoch, while in the right plot, they are currently decaying.}
\label{fish}
\end{figure}
For the analysis of the MCMC results, we adopt a frequentist approach so as to avoid the strong prior dependence induced by the fact that the posterior probability does not vanish at the edge of the prior range. Thus, we extract the $\chi^2$ from the MCMC sample, performing a profile likelihood analysis over the $\Omega_m$ and $\Omega_\Lambda$ degrees of freedom. We plot the iso-$\Delta \chi^2$ contours at $4.61$, which corresponds to a $90\%$~CL for the two remaining degrees of freedom, $m_\chi$ and $\st$. The regions excluded by our study are represented by the shaded areas in Fig.~\ref{H0clean}. The central light green region represents constraints arising from the HST prior and the Supernovae redshift-luminosity relationship, although we find that in practice, the latter do not contribute much to the exclusion. The purple region represents the effect of adding CMB and BAO constraints.  The right-hand sides of the green contour, representing lower bounds on the mass, roughly follows a line of equal lifetime. This essentially represents a limit on the amount of time an extra decaying matter component can be present in the early Universe before its effects are ruled out by observation. The HST prior on $H_0$ rules out lifetimes longer than about $\tau _s \simeq 10^{10}$\,s (roughly 300 years). Adding constraints from the epoch of recombination severely constraints the lifetime: anything above  $\tau_s \simeq 0.1$\,s will accelerate the early expansion rate enough to reduce the sound horizon at decoupling, shifting the BAO scale. However, if decay occurs before the BBN epoch and active neutrino decoupling, the decay products will just thermalize with the plasma, postponing the onset of BBN but leaving no further observational imprint. Thus, it is not possible to improve much further this lower bound on the sterile neutrino mass and mixing angle through cosmological observables, and we cut off our bound around $\tau_s \sim t_{\rm BBN}$. This is represented by the hatched region in Fig.~\ref{H0clean}. On the left-hand side of Fig.~\ref{H0clean}, constraints become weaker as the sterile neutrino becomes less massive. Even though the lifetimes are much longer than the age of the Universe, the extra particles essentially never change the overall equation of state enough to make a measurable difference. Lighter particles can be longer-lived, leading to an extra, small contribution to the overall radiation density, with very little effect on the background cosmology.

The thin blue subregion seen in Fig.~\ref{H0clean} represents the part of the parameter space for which the sterile neutrino can represent the cosmological dark matter. In this case, the matter component was constrained to be baryonic only, with $\Omega_m\equiv \Omega_b \simeq 0.04$. The slope closely follows the naive estimate outlined in Section \ref{sec:sterileTheory}, drawn as a red line labeled ``$\Omega_s h^2 \leq 0.12$''. The endpoint of this line is also readily understood: 
above 170 keV, the lifetime begins to approach the age of the Universe, and decays prevent the DM from dominating the matter component until today. With the data sets considered here, this region of points actually provide a good fit to the data, with a $\Delta \chi^2=1.6$ for the best-fit point along that line (relative to the absolute best-fit).  In practice, this scenario is severely constrained from the right by X-ray constraints on decaying DM (see \textit{e.g.} Ref.~\cite{Abazajian:2012ys}) and from the left by Lyman-$\alpha$ constraints on the matter power spectrum \cite{Viel:2013fqw}. These two constraints overlap, leaving no region in which the non-resonantly produced sterile neutrino can account for all of the dark matter.

We provide examples of  the relative contribution $\Omega_i \equiv \rho_i/\rho_c$ of each component to the total mass energy density of the Universe in Fig.~\ref{fish}, illustrating both \emph{allowed} and \emph{forbidden}  sterile neutrino cosmologies. 

The upper panel of Fig.~\ref{fish} shows the case for a heavy neutrino with a mass of $27$~keV and a tiny mixing with the light sector, $\st=10^{-10.2}$. This long-lived case corresponds to the Dodelson-Widrow keV sterile neutrino dark matter scenario, lying along the blue line in Fig.~\ref{H0clean}; this specific scenario is represented by a blue diamond.

The middle panels of Fig.~\ref{fish} illustrate scenarios ruled out by our cosmological data analyses.  On the left-hand side, we show the Universe's component evolution history for a heavy neutrino of mass $m_s=0.3$~MeV and with a mixing of $\st=10^{-4}$, shown by a green square in Fig.~\ref{H0clean}. In this region of parameter space there exists an extra period of matter domination in the early Universe, which can yield a much larger value of $H_0$, \textit{i.e.} a Universe that is younger than current cosmological observations indicate. Both the recombination and the drag redshift are much higher than in the $\Lambda$CDM standard picture, leading to a value of $r_s$ that is unreasonably small ($r_s \sim 1.2$~Mpc). 

On the right-hand middle panel of Fig.~\ref{fish}, we show the case of $m_s=100$~MeV and $\st=10^{-8}$, (light blue circle in  Fig.~\ref{H0clean}). This cosmology, due to an extended radiation epoch governed by the sterile neutrino decay products (which serves to compensate for the previous faster expansion) yields the correct age, and it is therefore allowed by HST and SNLS measurements. However, this type of ``conspiracy'' is ruled out once one considers the effect on recombination-era observables. By the time the optical depth of the Universe reaches $\tau = 1$, the Universe has been expanding more rapidly than in the standard $\Lambda\mathrm{CDM}$ scenario, restricting the distance traveled by acoustic perturbations ($r_s\sim 40$~Mpc in this  scenario). 

\begin{figure}[h]
\includegraphics[width=.7\textwidth]{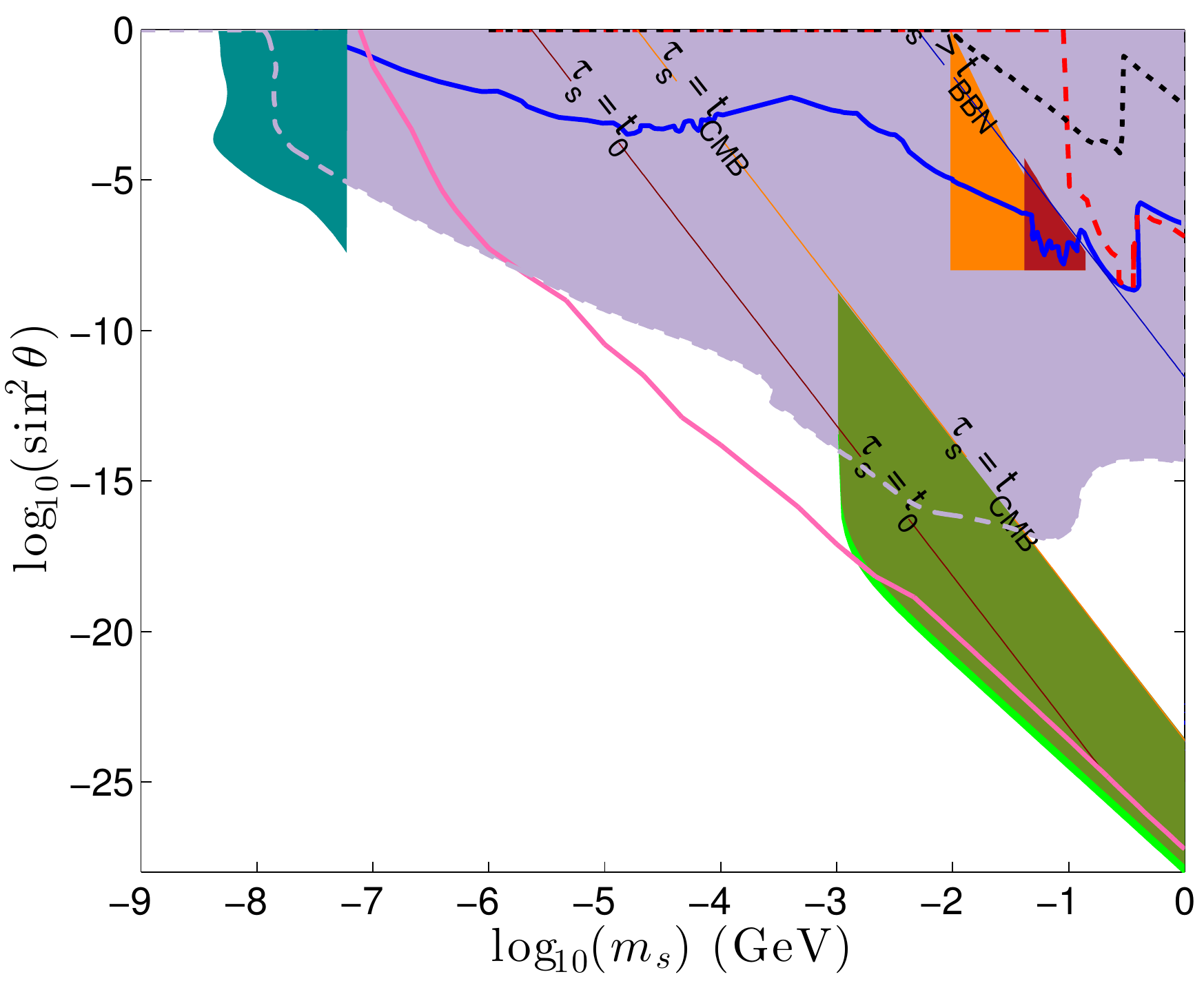}
\caption{Comparison of our cosmological bounds (purple shaded region) with exclusions from previous works. The lines represent upper limits from \cite{Atre:2009rg} on the mixing angle $\sin^2 \theta_{es}$ (solid blue),  $\sin^2 \theta_{\mu s}$ (dashed red) and  $\sin^2 \theta_{\tau s}$ (dotted black) based on accelerator data. The red and orange regions from \cite{Ruchayskiy2012} are excluded by constraints on helium production at BBN, as measured from astrophysics (orange) and CMB (red). The lower edge of these bands corresponds to the end of the range probed by Ref\,\cite{Ruchayskiy2012}. The green area is excluded from decays to electron-positron pairs, which raise the ionization floor after recombination and alter the CMB temperature and polarization power spectra \cite{Diamanti:2013bia}. Mixing only with $\nu_\mu$ or $\nu_\tau$ is assumed in the dark green area, whereas the excluded region extends down to the bright green area in the other extreme, if mixing is instead with $\nu_e$. The pink line denotes the upper limit based on X-ray constraints presented in \cite{Lattanzi:2013uza}, rescale to the relative abundance $\Omega_s/\Omega_{DM}$ at each point. Finally, the cyan region in the top-left corner corresponds to CMB limits on massive extra neutrinos in the linear regime.}
\label{fig:otherconstraints}
\end{figure}

The two lower panels of Fig.~\ref{fish} show two decaying scenarios allowed by the data considered here. However, in both cases, the population of the sterile states is non negligible but their decay occurs after recombination and therefore bounds from $\Delta N_{\textrm{eff}}$ at this epoch are avoided. The left panel (red circle in Fig.~\ref{H0clean}) depicts  the case in which the decay occurs before the current epoch. Notice that the sterile decay products can dominate the radiation component at late times. The presence of both the sterile and its nonthermal active neutrino decay products may affect standard structure formation; the precise calculation of this effect is beyond the scope of this study. We will see later that this scenario is furthermore excluded by CMB constraints (shown in Fig.\,\ref{fig:otherconstraints}). In the  bottom-right panel of Fig.~\ref{fish} (black star in Fig.~\ref{H0clean}), the sterile neutrino is currently decaying, producing a flux of $\sim 100$~keV neutrinos. Unfortunately, these energies are below the threshold of detectability for relic supernova neutrino searches~\cite{Bays:2011si}.

Electromagnetic energy injection into the intergalactic medium after recombination can also affect the observed CMB power spectrum ~\cite{Chen:2003gz,Padmanabhan:2005es,Zhang:2006fr,Zhang:2007zzh,Slatyer:2009yq,Chluba:2009uv,Galli:2009zc,Kanzaki:2009hf,Giesen:2012rp,Lopez-Honorez:2013cua,Diamanti:2013bia}, as well as its blackbody spectrum~\cite{Kawasaki1986,HuSilk1993PRD,Hu:1993gc,McDonald:2000bk,Chluba:2011hw}. Decays into electron-positron pairs raise the ionization floor of hydrogen and helium in the intergalactic medium, rescattering CMB photons. This suppresses correlations on small scales (large $\ell$) by smearing the last scattering surface, while boosting polarisation correlations on large scales (small $\ell$) by rescattering CMB light at low redshift. CMB observations and Lyman alpha measurements of the IGM temperature constrain dark matter decays into $e^+e^-$ pairs, yielding the limit $\tau_{DM} \gtrsim 4 \times 10^{25}$s \cite{Diamanti:2013bia}. This can be translated into a bound on the sterile neutrino lifetime via:
\begin{equation}
b_e \frac{\Omega_s}{\Omega_{DM}} \Gamma_s \leq (4 \times 10^{25} \mathrm{s})^{-1},
\end{equation}
where $b_e$ is the ratio of decays into $e^+e^-$ versus $3\nu$. For $m_s > 2m_e$, this is approximately the line
\begin{equation}
\sin^2 \theta \gtrsim 3 \times 10^{-4} \sqrt{ \frac{192 \pi^3}{b_e}\left(\frac{\mathrm{GeV}^{-2}}{ G_F}\right)^2\left(\frac{\mathrm{eV}}{m_s}\right)^7}.
\end{equation}
The ratio $b_e$ is computed via the decay rates given of Ref.~\cite{Gorbunov:2007ak}. If $\tau_s \lesssim t_{\rm CMB}$, this bound does not apply, since the resulting decay products will recombine with the rest of the plasma, although decays that occur very shortly before recombination can affect the blackbody spectrum of the CMB; such bounds have not been computed as rigorously as in the $\tau_s < t_{\rm CMB}$ case.

In Fig.~\ref{fig:otherconstraints} we compare our cosmological bounds with the above constraints. The blue, red and black line respectively show accelerator upper limits on the sterile mixing with electron, muon and tau neutrinos from Ref.~\cite{Atre:2009rg}. The orange and red regions show constraints from sterile neutrino decays during BBN, which would affect the primordial helium abundance; finally, the green region in the lower part of the figure shows the CMB bound on a dark species decaying to electron-positron pairs. Mixing only with $\nu_\mu$ or $\nu_\tau$ is assumed in the dark green area, whereas the excluded region extends down to the bright green area if mixing is instead with $\nu_e$. Any other combination of mixing ratios would yield a bound inside that sliver.

Fig.~\ref{fig:otherconstraints} also shows the full CMB constraints on massive extra neutrino species, in dark cyan in the upper left-hand corner. We computed these using the CAMB and \texttt{CosmoMC} cosmology packages  \cite{Lewis:2002ah,Lewis:1999bs,Howlett:2012mh,Lewis:2013hha}, with data from Planck \cite{Ade:2013lta,Planck:2013kta,Ade:2013mta}, WMAP9 \cite{Bennett:2012zja},  together with the BAO constraints previously detailed and a HST prior on the Hubble constant. These constraints are stronger than our background-only limits, but we cut them off around 60 eV, when $\nu_s$ begins to behave as warm dark matter, beyond which an in-depth treatment of the perturbation theory at highly non-linear scales would be required.

Finally, X-ray bounds strongly constrain sterile neutrino dark matter. We approximate these using the limits collected in Ref. \cite{Lattanzi:2013uza}.
While these were presented as bounds on Majoron dark matter decay, they can be rescaled to apply equally well to any cosmological species that decays to monoenergetic photons. In particular, the upper limits in Ref. \cite{Lattanzi:2013uza} should be relaxed by a factor 2, to account for the fact that Majorons decay to two photons, while sterile neutrino decay produces a single photon. We further rescale these limits by a factor $\Omega_s/\Omega_m$ as given in Eq. (\ref{OmegaBound}), to account for the dark matter fraction made up by sterile neutrinos. In principle, one should also take into account that a given photon energy could correspond to a different mass of the parent particle in the two models. However, there is no need for this in this particular case, since
the photon(s) produced in both Majoron and sterile neutrino decay carry an energy equal to half the mass of the parent particle. The X-ray upper limit on $\sin^2\theta$ is shown as a pink line in Fig. \ref{fig:otherconstraints}. It should be kept in mind that this limit does not extend far beyond the $\tau = t_0$ line: sterile neutrinos that decay fast enough will not remain to produce X-ray signals today. Our background cosmology constraint thus fills an important gap between the X-ray and collider lines in the heavy mass ($m_s \simeq 1$ GeV) region, as well as filling the gap between the CMB limits on the mass of light (i.e., acting as hot dark matter) neutrino species and the X-ray limits on decaying dark matter.

\subsubsection{Decays to heavier species}
\begin{figure}[h]
\includegraphics[width=.7\textwidth]{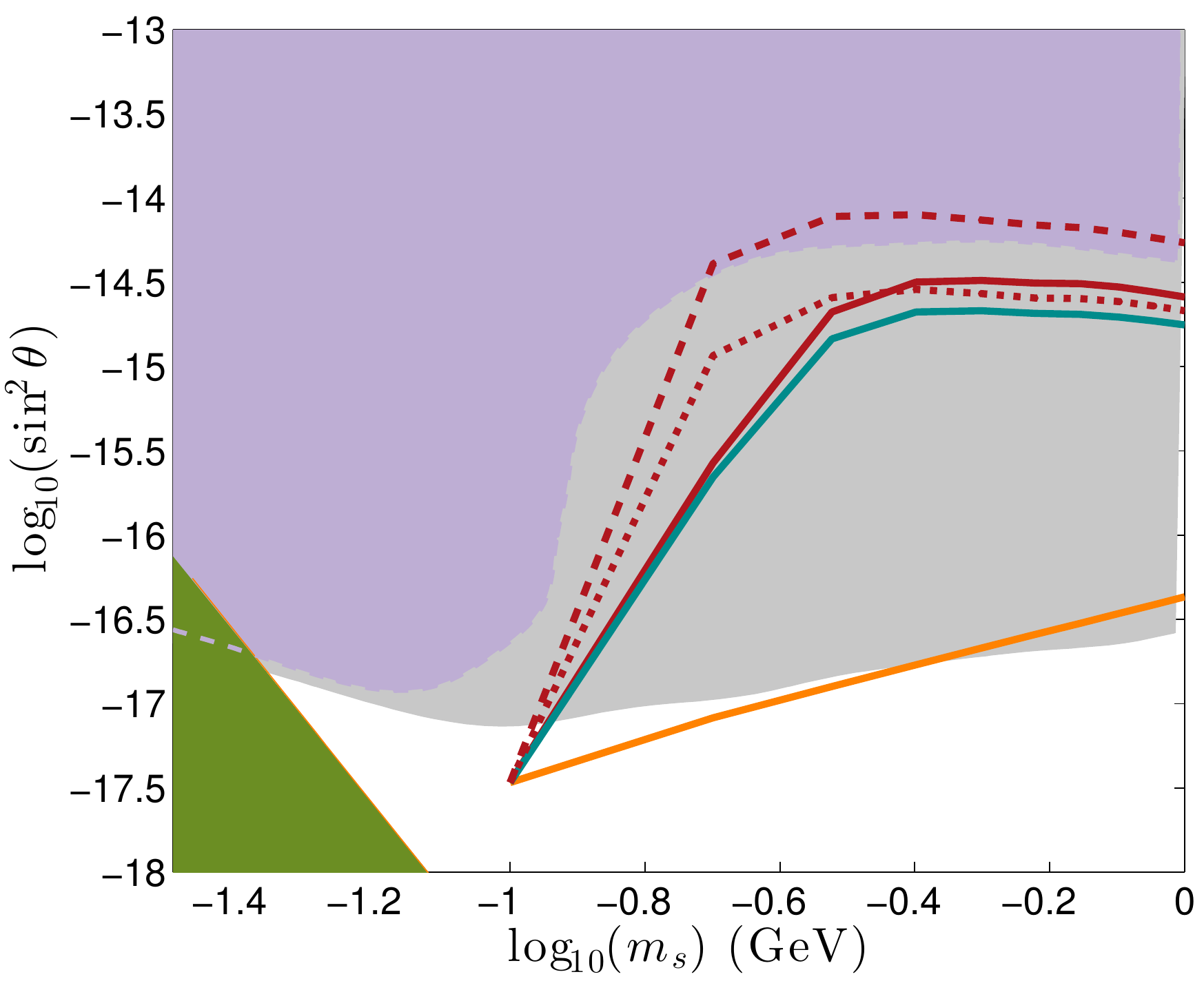}
\caption{Close-up of the lower right-hand corner of Fig. \ref{fig:otherconstraints}, illustrating the relaxation of cosmological constraints of sterile neutrinos from including decays to heavier species. The straight orange line shows the $\Delta N_{eff} = 1$ line when only decays to $3\nu$ are considered, as in our Monte Carlo (light grey region). The solid purple region corresponds to our MCMC constraints assuming coupling to $\nu_e$; these do not include reheating of the plasma. The red lines show the relaxation of this constraint when decays to heavier species are included: long dashed lines include coupling to electron neutrinos only, solid red to $\nu_\mu$, and the intermediate short-dashed line corresponds to democratic coupling to all three flavours. The  cyan line below these curves illustrates the muon neutrino case without dilution of the plasma taken into account. }
\label{fig:zoomin}
\end{figure}

When the mass $m_s$ reaches $m_\mu$, decays to heavier species are possible.  The impact of this type of effect in Big Bang Nuclosynthesis (BBN) was considered in full detail in Ref.~\cite{Ruchayskiy2012} for a range of masses $10-140$ MeV,  see also the works of Refs.~\cite{Dolgov:1987qz,Kawasaki:1992kg,Dolgov:1997it,Kawasaki:2000en,Dolgov:2000pj,Hannestad:2004px,Smith:2008ic}. This can have a large effect on the decay rate, as the branching ratio to $3\nu$ becomes subdominant. This leads to two effects: first, a relaxation of constraints due to faster decay into light species; second, an injection of heat into the plasma due to decay into electromagnetically interacting particles. The light region in the lower-right of Fig.~\ref{H0clean} shows the first of these effects. Since it would be too time-consuming\footnote{Specifically, the parametrization of the radiation density in Appendix \ref{cosmoappendix} would no longer apply.} to include the extra energy injection in our Monte Carlos, we illustrate the second effect in Fig.~\ref{fig:zoomin}, showing the modification to the $\Delta N_{eff} = 1$ line. We include decays to $\pi^0 \nu$, $\pi^\pm e^\mp$, $\pi^\pm \mu^\mp$, $K^\pm e^\mp$ and the three-body decays to $\nu e^+ e^-$ and $\nu \mu^+ \mu^-$, using the branching ratios given in \cite{Gorbunov:2007ak}. We show three cases: interactions only with $\nu_e$, only with $\nu_\mu$, and democratic to all three flavours. We also show an example in dark cyan (coupling to $\nu_\mu$) where the branching to all channels is included, but energy injection from charged particles is neglected. 

\section{Conclusions}
\label{sec:discussion}
The most natural extension of the Standard Model (SM) of particle physics able to accommodate neutrino masses and mixings implies the existence of extra sterile neutrino states at a new mass scale to be determined observationally. In this work we study the impact of such extra sterile neutrinos on cosmological measurements related exclusively to the Universe's background expansion history. In particular we used measurements of the Hubble constant, Supernovae Ia luminosity distances, the Cosmic Microwave Background shift parameter, as well as measurements of the Baryon Acoustic Oscillation scale to set bounds on the mixings of a general extra sterile neutrino species with the SM active flavours as a function of the sterile neutrino mass. We find that, for the broad range of sterile neutrino masses explored here, ranging from eV's up to the GeV region, the impact of the sterile neutrino on the background cosmology is very significant. Indeed, even assuming very small mixing angles,  the relic density of sterile neutrinos will be set while still relativistic, avoiding therefore Boltzmann suppression, and ensuring a non negligible number density. However, as the Universe expands and cools, they will become non-relativistic and thus their relative contribution to the total energy density is enhanced. This allows to derive very competitive constraints on their allowed parameter space.  Two main bounds can be derived. In the high mass region, where the sterile neutrinos are short lived, we find that for a sterile neutrino decaying after BBN and before recombination, { only sterile-active mixing angles  up to $\sin^2 \theta\lesssim10^{-16}$ below the muon production threshold or up to $\sin^2 \theta\lesssim 10^{-14}$ above it, are allowed at $90\%$~CL. There is a sharp relaxation of the limit around the muon threshold, mostly due to the faster decay into light species. After recombination and  above the $e^+e^-$ production threshold, CMB constraints on decaying matter are much stronger ranging from $\sin^2 \theta\lesssim10^{-17}-10^{-27}$ for masses
between 1 MeV and 1 GeV. Since these bounds quickly degrade for shorter lifetimes a significant region of small mixing angles below $10^{-14}$ remains 
unconstrained, bounded from above by background cosmology and from below by $e^+e^-$ CMB constraints in the range between 100~MeV and 1~GeV. 

In the low mass region, where the sterile neutrinos are long-lived, the current cosmological amount of dark matter sets  an \emph{upper} bound on the sterile neutrino mixing. Below $m_s = 100$\, MeV, this leads to the rather strong condition: 
\begin{equation}
\sin^2\theta \lesssim 0.026(m_s/\mathrm{eV})^{-2}  \, \, \, \, \, \,\, \, \, (m_s \lesssim 100 \, \mathrm{MeV})
\end{equation}
at 90\% CL. This bound is improved by X-ray bounds in the range from 1KeV-1MeV. Below 60 eV, we have added the full CMB constraints including perturbations. The bounds disappear for masses below a few eV.

The cosmological bounds improve significantly over laboratory searches in the same mass range~\cite{Atre:2009rg,Ruchayskiy:2011aa}. For the lightest masses the dominant laboratory constraints are on the mixing of the sterile neutrino with electrons through kink searches in $\beta$-decays (up to $m_s \sim 100$~keV) and peak searches in pion or kaon decays at higher masses. The constraints range from $\st < 10^{-2}$ to $\st < 10^{-5}$ for $m_s$ between 1~keV and 10~MeV, to be compared with $\st < 10^{-8}$ and $\st < 10^{-16}$ from the impact of the sterile neutrino on the background cosmology. The laboratory bounds can be improved by 1 or 2 orders of magnitude from the constraints from neutrinoless double beta decay in the case of Majorana neutrinos. However, for sterile neutrinos lighter than $\sim 100$~MeV a cancellation between their contribution and the SM neutrino contributions is usually present, relaxing this constraint and rendering it quite model-dependent~\cite{Blennow:2010th}. Given their higher masses, it is difficult to constrain the mixing of sterile neutrinos in this mass range with muons and taus. For masses between the MeV and GeV range, laboratory searches yield stronger constraints. Indeed, peak searches in meson decays as well as searches for sterile neutrino decays at colliders translate in constraints between $\st < 10^{-4}$ to $\st < 10^{-8}$ in the mass range between 10~MeV and 1~GeV for mixings with electrons and muons and $\st < 10^{-2}$ to $\st < 10^{-4}$ for mixing with taus. 

 Taken together, our results show that for a large range of masses, current cosmological measurements lead to limits on the active-heavy neutrino mixing that are around 10 orders of magnitude stronger than the direct laboratory searches, thus illustrating that cosmology remains a powerful complementary tool in the search for particle physics beyond the Standard Model.

\section*{Acknowledgements}
We thank Sergio Palomares-Ruiz for his valuable comments and suggestions. OM is supported by the Consolider Ingenio project CSD2007--00060, by
PROMETEO/2009/116, by the Spanish Grant FPA2011--29678 of the MINECO. 
ACV is supported by FQRNT and European contract FP7-PEOPLE-2011-ITN.  
EFM acknowledges financial support by the European Union through the FP7 Marie Curie Actions CIG NeuProbes (PCIG11-GA-2012-321582) and the Spanish MINECO through the ``Ram\'on y Cajal" programme (RYC2011-07710) and through the project FPA2009-09017.
The authors are also partially supported by PITN-GA-2011-289442-INVISIBLES. 
We also thank the Spanish MINECO (Centro de excelencia Severo Ochoa Program) under grant SEV-2012-0249.
Part of this work was carried out
while M.L. was visiting the Instituto de F\'isica Corpuscular in Valencia, whose hospitality is
kindly acknowledged, supported by the grant \emph{Giovani ricercatori} of the University of Ferrara,
financed through the funds \emph{Fondi 5x1000 Anno 2010} and \emph{Fondi Unicredit
2013}.

\appendix
\section{The background cosmology}
\label{cosmoappendix}
In principle, one can write the evolution equation of each species $i$ which contributes to the Friedmann equation (\ref{friedmann}) as:
\begin{equation}
y_i'(x)= -3(1+w_i)y_i(x); 
\label{ystandard} 
\end{equation}
$i$ in (\ref{ystandard}) represents radiation and standard neutrinos (which evolve together with $w_r = 1/3$), matter ($w_m = 0$) and the cosmological constant ($w_\Lambda = -1$). Once again, $x \equiv \ln a$ is the log of the scale factor, and $y_i \equiv \rho/\tilde \rho$ is the dimensionless energy density. We neglect the active neutrino masses, as they only become important at very late times, once the neutrinos contribute only $\sim 10^{-5}$ to the total density. The solutions of (\ref{ystandard}) for matter and $\Lambda$ are quite trivial: $y_m(a) = y_m(1)a^{-3}$ and $y_\Lambda(a) = y_\Lambda(1)$.  When heavy species fall out of equilibrium with radiation, however, one must be careful to account for the change in relativistic degrees of freedom, leading to slight deviations of the radiation component from (\ref{ystandard}). Instead, we write the radiation density as a function of temperature:
\begin{equation}
\rho_{rad}(T_{\rm rad}) = \frac{\pi^2}{30}g_*(T_{\rm rad}) T_{\rm rad}^4, 
\label{rhoofT}
\end{equation}
where $g_*$ is the usual effective relativistic number of degrees of freedom. From the conservation of entropy it can be written in terms of the scale factor $a$ by inverting:
\begin{equation}
a(T_{\rm rad}) = \left(\frac{g_{*s}(T_0)}{g_{*s}(T_{\rm rad})}\right)^{1/3}\frac{T_0}{T_{\rm rad}}.
\label{aofT}
\end{equation}
$g_{*}(T)$ and $g_{*s}(T)$ can be written in terms of the parametrization of \cite{2010PhRvD..82l3508W}. Analogously, it is convenient to parametrize $\rho_{\rm rad}(x)$ obtained from (\ref{rhoofT}) and (\ref{aofT}), where, again, $x = \ln (a)$. We find:
\begin{equation}
\rho_{\rm rad}(x)  = (10^{-51} \, \, \mathrm{GeV}^4) \times e^{-4 x} \left(a_0 +  \sum_{i = 1}^{5} a_{i} \left[ 1 + \tanh\left(\frac{x+b_i}{c_i}\right) \right]\right),
\label{rhofit}
\end{equation}
where $a_0 = 1.233$ and the other parameters $a,b,c_{i= 1...5}$ are given in Tab. \ref{rhotable}.
\begin{table}[h]
\caption{Coefficients for the fit to $\rho_{\rm rad} (a)$ given in Eq. (\ref{rhofit})}
\begin{tabular}{c | c c c c c}
 $i$& 1 & 2 & 3 & 4 & 5 \\ \hline
 $a_i$ & 0.2832 & 0.2431 & 0.3535 & 0.9527 & -0.7708 \\
 $b_i$ &  20.71 & 26.71 & 28.00 & 32.10 & 32.15 \\
 $c_i$ & 0.8594 & 1.044 & 0.2168 & 2.375 &  2.013 
\end{tabular}
\label{rhotable}
\end{table}

\bibliographystyle{JHEP_mod}
\bibliography{Neff}

\end{document}